\documentclass[useAMS,usenatbib,usegraphicx]{mn2e}
\usepackage{amsmath,amsfonts,amssymb}

\newcommand{\hm}{\mbox{$h^{-1}$}}

\newcommand{\bj}{\mbox{$b_{\rm J}$}}
\newcommand{\1}{^{-1}}
\newcommand{\paperone}{Paper~1}
\def\Msun{\hbox{$\rm\, M_{\odot}$}}

\newcommand{\changed}[1]{{\bf #1}}

\title [Tidal Disruption of Satellite Galaxies]{Tidal Disruption of
  Satellite Galaxies in a Semi-Analytic Model of Galaxy Formation}

\author[Henriques et al.]
{Bruno M. B. Henriques$^{1,2}$\thanks{E-mail: Bruno.Henriques@port.ac.uk},
 Peter A. Thomas$^{2}$ \\
 {}$^{1}$Institute of Cosmology and Gravitation, University of
 Portsmouth, Portsmouth PO1 3FX, United Kingdom\\
 {}$^{2}$Astronomy Centre, University of Sussex, Falmer, Brighton BN1 9QH,
 United Kingdom}

\begin{document}

\date{Submitted to MNRAS}

\pagerange{\pageref{firstpage}--\pageref{lastpage}} \pubyear{2009}

\maketitle

\label{firstpage}

\begin{abstract}

  We introduce a new physical recipe into the De Lucia and Blaizot
  version of the Munich semi-analytic model built upon the
  Millennium dark matter simulation: the tidal stripping of stellar
  material from satellite galaxies during mergers.

  To test the significance of the new physical process we apply a
  Monte Carlo Markov Chain parameter estimation technique constraining
  the model with the $K$-band luminosity function, $B-V$ colours and the
  black hole-bulge mass relation. The differences in parameter
  correlations, and in the allowed regions in likelihood space, reveal
  the impact of the new physics on the basic ingredients of the model,
  such as the star-formation laws, feedback recipes and the black hole
  growth model.

  With satellite disruption in place, we get a model likelihood four
  times higher than in the original model, indicating that the new
  process seems to be favoured by observations. This is achieved
  mainly due to a reduction in black hole growth that produces a
  better agreement between the properties of central black holes and
  host galaxies. Compared to the best-fit model without disruption,
  the new model removes the excess of dwarf galaxies in the original
  recipe with a more modest supernova heating.

  The new model is now consistent with the three observational data
  sets used to constrain it, while significantly improving the agreement with
  observations for the distribution of metals in stars. Moreover, the
  model now has predictions for the intra-cluster light, a very
  significant component of large groups and clusters, that agree with
  observational estimates.

\end{abstract}


\section{Introduction}
\label{sec:intro}

In recent years, semi-analytic (SA) models have experienced a
significant degree of success, achieving a considerable agreement with
a large set of observational properties. These range from the
luminosity to the stellar mass functions, Tully-Fisher relations,
clustering, galaxy colours, etc \citep{Bower2006, Croton2006}.

Some fundamental changes determined the improvements achieved.  First
of all, these recipes were implemented on top of a direct dark matter
simulation, which just recently have been able to simulate
cosmological volumes with a large enough resolution to follow haloes
containing dwarf galaxies \citep[the Millennium
run,][]{Springel2005}. Secondly, some key physical recipes were
introduced: the energy feedback from supernovae explosions, which for
small enough galaxies and strong enough explosions can drive the gas
out of the galaxy \citep{Benson2003, Delucia2004}; and the feedback
from central black holes, which determines the properties of massive
galaxies \citep{Granato2004, Bower2006, Cattaneo2006, Croton2006,
  Menci2006, Monaco2007, Somerville2008}.

The level of agreement achieved opens up a number of
possibilities. This agreement can now be quantified
against different observational data sets using robust statistical
tests (\citealt{Kampakoglou2008}; \citealt[hereafter
  \paperone]{Henriques2009}). This means that the allowed likelihood
regions in parameter space can be obtained and confidence limits for
the preferred parameter values can be built. Moreover, with the
physics that determine the global properties of galaxies reasonably
well understood, modellers can now focus on additional ingredients
that determine the fine tuning of galaxies properties.

Amongst these detailed studies we can find the stripping of gas from
satellite galaxies \citep{Font2008}; the impact of the assumed dust
model on the overall galaxy properties \citep{Delucia2007} and on the
galaxy clustering over redshift \citep{Guo2009}; the impact of quasar
mode feedback on the luminosity-temperature (L-T) relation in clusters
\citep{Bower2008}; and the effects of a dynamical treatment of
galactic winds on host galaxies \citep{Bertone2007}.

Another physical process that is becoming more relevant in theoretical
models is the disruption of stellar material from merging satellites
due to tidal stripping, with studies being performed using both the
semi-analytic \citep{Bullock2001, Taylor2001, Benson2002b, Monaco2007,
  Henriques2008, Somerville2008, Kim2009} and the Halo Occupation
Distribution (HOD) approach \citep{Yang2009, Wetzel2009} .

Evidence for the importance of this process in galaxy formation comes from
various fields. The existence of a diffuse population of intra-cluster stars was first
proposed by \citet{zwicky1951} and has since been detected
unambiguously \citep{durrell2002, Gal-Yam2003, Neill2005, Krick2006}.
The light associated with intra-cluster stars, or diffuse intra-cluster
light (ICL), can contribute between 10 and 40 per cent of the optical
emission of rich galaxy groups and clusters (\citealt{Bernstein1995};
\citealt{Gonzalez2000}; \citealt{Feldmeier2002};
\citealt{Feldmeier2004b}; \citealt{Gonzalez2005};
\citealt{Zibetti2005}).

Rather than having formed in the intra-cluster medium, gas-dynamical
simulations generally agree that the bulk of the ICL is emitted by
stars that have been continually stripped from member galaxies
throughout the lifetime of a cluster, or have been ejected into
intergalactic space by merging galaxy groups \citep{Moore1996, 
Napolitano2003, Murante2004, Willman2004, Sommer-Larsen2005, 
Monaco2006, Murante2007, Rudick2009}.

Observationally, low surface brightness features have been identified
in the Coma and Centaurus clusters (\citealt{Gregg1998};
\citealt{Trentham1998}; \citealt{Feldmeier2002}), indicating the
presence of dynamically-young tidal structures produced by the
disruption of infalling galaxies. Moreover, \citet{Faltenbacher2005}
show that the projected number density profile of dwarf galaxies in
NGC 5044 can only be explained by assuming that a significant amount
of mass in satellite galaxies is tidally disrupted.

If, as the evidence suggests, intra-cluster stars are the remnants of
material stripped from merging satellites, the build up of this
component can be directly followed by a semi-analytic model with a
self consistent implementation of tidal disruption \citep{Bullock2001,
  Taylor2001, Benson2002a, Monaco2006, Somerville2008}. Also,
the disruption of material from satellite galaxies might play a
crucial role in solving a common problem to most semi-analytic models:
the excess of dwarf galaxies \citep{Henriques2008,
  Weinmann2006b}.

Despite the fact that a number of authors have introduced satellite
disruption in theoretical models of galaxy formation, the complexity
of these models makes it difficult to fully understand the impact of
different physics on the global galaxy properties. Also, the number of
free parameters makes it some times impossible to take full advantage
of the new physics introduced, when a manual tuning fails to reach the
best possible fit.

In this paper we make use of the Monte Carlo Markov Chain (MCMC)
statistical techniques introduced in \citet{Henriques2009} to test the
impact of satellite disruption in semi-analytic models of galaxy
formation. With the introduction of a new physical recipe, both the
normalisation and shapes of the acceptable likelihood regions contain
useful information that can be used to discriminate between models.
We will show that the new model is favored by the data, that it
provides a better fit to the metallicity of galaxies than our previous
model, and that it predicts ICL fractions that agree with
observations. We note that similar MCMC techniques were proposed
independently by \citet{Kampakoglou2008} to test the impact of
different star formation modes in their specific semi-analytic recipe.

This paper is organized as follows. In Section 2, we describe the
original semi-analytic model used in our study and the new
implementation of satellite disruption. Section 3 presents results
from a model with disruption using the same parameters as in the
original recipe to directly test the differences between the two. In
Section 4 we briefly describe the MCMC techniques introduced in
\paperone. In Section 5 we present the best fit model with satellite
disruption and compared its likelihood maximum, acceptable likelihood
regions and galaxy predictions with the previous model. Finally in
Section 6 we summarize our conclusions.

\section{The model}
\label{sec:model}

To perform the work described in this paper we used the Munich SA
recipe as described in \citet{Croton2006} and \citet[][hereafter
  DLB07]{Delucia2007}.  It is built on top of a direct numerical
simulations of the dark matter structure in a cosmological volume, the
Millennium Simulation \citep{Springel2005}. This simulation traces the
evolution of dark matter haloes in a cubic box of 500\hm Mpc on a
side. It assumes a $\Lambda$CDM cosmology with parameters $\Omega_{\rm
  m}=0.25$, $\Omega_{\rm b}=0.045$, $h=0.73$, $\Omega_\Lambda=0.75$,
$n=1$, and $\sigma_8=0.9$, where the Hubble parameter is $H_0 = 100$
\hm km s$\1$ Mpc$\1$. The simulation follows $2160^3$ dark matter
particles of mass $8.6\times 10^8$ \hm M$_{\sun}$. Since dark matter
haloes are required to contain at least 20 particles, the minimum halo
mass is $1.7\times 10^{10}$ \hm M$_{\sun}$, with a corresponding
baryonic mass of about $3.1\times10^9$ \hm M$_{\sun}$.

The galaxy formation model itself follows the evolution of baryons
from when they collapse into a hot gas phase, through cooling onto a
disk where stars can form. As the most massive stars die, supernovae
(SN) eject energy into the surrounding medium, reheating the cold gas
back into the hot phase or even ejecting it into the external
reservoir. The black hole evolution is modeled and in massive galaxies
the mechanical heating from its quiescent growth suppresses the
cooling. Mergers generate star formation bursts and, depending on the
mass ratio between the galaxies, disks are destroyed to form
bulges. Finally dust and stellar population synthesis models transform
the predicted quantities into galaxy properties that can be compared
with observations.

With all the recipes in place, the model is able to predict reasonably
well the shape of the luminosity function in different bands, the
stellar mass function, the galaxy colours, the black hole-bulge mass
relation, the metallicity of stars, the Tully-Fisher relation, etc.
However some challenges remain. For example, as happens with most
current SA recipes, this model has known problems in
correctly predicting the properties of dwarf galaxies
\citep{Croton2006, Weinmann2006b, Henriques2008, Henriques2009} and it
has no predictions for the intra-cluster light, a component that
can be substantial for the largest groups in the local Universe.

These two aspects lead us to introduce a new physical recipe: the
stripping of stellar material from satellite galaxies due to the tidal
forces they experience from the central companions. By removing this
material from merging galaxies we expect to decrease their size and
hence the number density of low-mass galaxies (where an excess is seen
in the original model). At the same time, by moving the disrupted
material into the intra-cluster medium (ICM) we will be able to predict the
properties of the ICL. Moreover, since this corresponds to stars that
would otherwise end up in the central galaxies of large groups and
clusters, we expect to see considerable changes in the properties of
the most massive objects in the model.

To introduce the new physics we use the implementation proposed by
\citet{Taylor2001} and follow the radial position of satellites as
they merge into central objects. At each location, we compute the
radius relative to the satellite centre at which the sum of the tidal
force from the parent halo and the centrifugal force equal the
self-gravitational force of the satellite.  Any material outside this
radius at each time step is considered to be unbound and is moved from
the satellite galaxy into the ICM.

 We assume that disruption of stellar material only acts on
  galaxies that have experienced large enough tidal forces to
  completely strip their dark matter component (for those who are
  aware of the Munich SA terminology, type 2 galaxies).

\subsection{Dynamical Friction}
The original model of DLB07 follows the merging of satellites by simply
setting a up a ``merging clock'', whenever a satellite loses its dark
matter halo.  After this time is elapsed the galaxy is assumed to merge
with its central companion. 

The merging time is calculated using the \citet{Chandrasekhar1943} formula
for the dynamical friction force acting on the satellite, as described in \citet{Binney1987}:
\begin{equation} \label{eq:msamdf1}
F_{\rm{df}}=-\frac{4\pi {\rm{G}}^2 m^2_{\rm{sat}} \ln(\Lambda) \rho
  B(x)}{v^2_{\rm{sat}}},
\end{equation}
where $m_{\rm{sat}}$ represents the satellite mass, $\ln(\Lambda)$ is
the Coulomb logarithm (assumed to be
$\ln\left(1+m_{\rm halo}/m_{\rm{sat}}\right)$), $\rho$ is the local density and
$v_{\rm{sat}}$ represents the orbital velocity of the
satellite within the halo.  $B(x)$ is the error function and we take
$x=|v_{\rm{sat}}|/\sqrt{2}\sigma_{\rm{halo}}=1$, where $\sigma_{\rm{halo}}$
represents the 1-D velocity dispersion of the halo.

Assuming circular orbits and an isothermal distribution of the total
mass in haloes, the previous equation can be integrated to give the
merging time:
\begin{equation} \label{eq:msamclock}
t_{\rm df}\approx1.17\frac{V_{\rm{vir}}r^2_{\rm{sat}}}
{{\rm{G}}\,m_{\rm{sat}}\ln(\Lambda)},
\end{equation}
where $r_{\rm{sat}}$ is the halocentric radius of the satellite at the
time that it lost its dark matter halo and $V_\mathrm{vir}$ is the virial
velocity of the central halo.

For our implementation we use the same equation, but we now record the
orbital radius of the satellite as it falls towards the halo centre.
In this way, at each time step, we can compute the forces acting on
the satellite at that radius and determine the amount of stellar
material that becomes unbound.

We note that the drag force acting on satellites and causing them to
merge with central objects depends on their mass. Because of this,
despite us using the same equation for the dynamical friction as in
DLB07, the implementation of satellite mass loss will change the
predicted merging times. As satellites lose their stellar material
the drag force they experience will decrease, causing them to survive
longer before being accreted by the central galaxy.

\subsection{Tidal Disruption}
\label{subsec:disdisruption}

The mass loss on satellites occurs through the action of tidal
forces. Assuming a slowly-varying system (a satellite in a circular
orbit) with a spherically-symmetric mass distribution, material
outside the tidal radius will be stripped from the satellite. This
radius can be identified as the distance from the satellite centre at
which the radial forces acting on it cancel out \citep{King1962,
  Binney1987}.  These forces are the gravitational binding force of
the satellite, the tidal force from the central halo and the
centrifugal force. Using the \citet{King1962} model and as described
in \citet{Taylor2001} the disruption radius $r_{\rm{t}}$ will be given
by:
\begin{equation} \label{eq:disdisruption1}
r_{\rm t} \approx \left(\frac{{\rm{G}}\,m_{\rm{sat}}}{\omega_{\rm{sat}}^2-\rm{d}^2\phi/\rm{d}r^2}\right)^{1/3},
\end{equation}
where $\omega_{\rm sat}$ is the orbital angular velocity of the satellite
and $\phi$ represents the potential of the halo.

Using the isothermal sphere approximation for the mass distribution of
the central halo and satellite galaxy, and assuming that the satellite follows
a circular orbit, equation \ref{eq:disdisruption1} can be rewritten as:
\begin{equation} \label{eq:disdisruption2}
r_{\rm t} \approx
\frac{1}{\sqrt{2}}\,\frac{\sigma_{\rm{sat}}}{\sigma_{\rm halo}}\,r_{\rm sat},
\end{equation}
where $\sigma_{\rm sat}$ and $\sigma_{\rm halo}$ are the velocity
dispersions of the satellite and the halo, respectively, the former
being estimated just before it becomes stripped.

The material outside this radius is assumed to be disrupted and
becomes part of the ICM. We assume that the galaxy as a uniform
metallicity distribution so that equal fractions of stellar mass and
metals are stripped away from the galaxy. We keep following the
satellites until their orbital radius becomes smaller than the sum of
the radii of the central and satellite galaxies, at which point we
assume a merger has occurred.

To compute the forces acting on satellite galaxies, we assume that the
total mass follows a spherically-symmetric isothermal sphere. However,
to determine the amount of material outside the disruption radius that
is lost at each time step, we use appropriate distribution models for
the stellar mass in the different galaxy components. The model
naturally defines galaxies as a combination of a disc and a
spheroid. We now explain how these two components are modelled.

\subsubsection{Mass Distribution in Disks}
We model the disc component as an exponential mass distribution given by:
\begin{equation} \label{eq:disdisk1}
M_{\rm{disk}}(<R)=\int_0^R2\pi R\,\Sigma_{d,0}\exp^{-R/R_d}dR,
\end{equation}
where $\Sigma_{d,0}$ gives the surface density of the disk and
the disk scale-length ($R_{\rm d}$) is calculated using the
formalism derived by \citet{Mo1998}. Integrating equation
\ref{eq:disdisk1}, the disk mass inside a given radius R is given by:
\begin{equation} \label{eq:disdisk2}
M_{\rm{disk}}(<R)=M_{\rm{disk}}
\left[1-\left(1+\frac{R}{R_{\rm d}}\right)\exp^{-R/R_{\rm d}}\right].
\end{equation}

\subsubsection{Mass Distribution in Bulges}
The mass in bulges is spherically symmetric distributed according to an
$r^{1/4}$ law:
\begin{equation} \label{eq:disbulge1}
\Sigma_{\rm{bulge}}=\Sigma_{b,0} 
\exp^{-7.67\left[\left(\frac{R}{R_b}\right)^{\frac{1}{4}}-1\right]},
\end{equation}
where $\Sigma_{b,0}$ gives the surface density of the bulge at the
effective radius, $R_b$, that contains half the projected light.
However, the integral to turn this into mass as a function of radius
has no analytic solution and so we use the following approximation instead:
\begin{equation} \label{eq:disbulge2}
M_{\rm{bulge}}(<r)=M_{\rm{bulge}}\int_0^{r/a}\frac{x^2}{x(1+x)^3}dx,
\end{equation}
where $M_{\rm{bulge}}$ is the total stellar mass of the bulge and the auxiliary
variable $x=R/a$ where $a=0.56R_b$. Integrating equation
\ref{eq:disbulge2}, the bulge mass inside a given radius r is given by
\begin{equation} \label{eq:disbulge3}
M_{\rm{bulge}}(<r)=M_{\rm{bulge}}\,\frac{r^2}{r^2+a^2}.
\end{equation}

The pioneering work of \citet{Kormendy1977} demonstrated that there is
a correlation between the effective radius ($R_b$) and the effective
surface brightness ($I$) of ellipticals galaxies. At the same time,
\citet{Faber1976} showed that the luminosity of these galaxies is
proportional to the velocity dispersion, with the derived relation
$L\propto\sigma^4$ expressing the intuitive notion that more luminous
galaxies have higher velocity dispersions (and hence higher
masses). From these two correlations, and since $L=\pi R_b^2 \langle
I\rangle$, it follows that the effective radius $R_b$ should be
correlated with the velocity dispersion $\sigma$. Since the velocity
dispersion is an intrinsic quantity of theoretical galaxies derived
from the properties of their host halos, we can use it to determine
the effective radius and the mass distribution of bulges. To do so, we
use the proportionality law from \citet{Djorgovski1987}:

\begin{equation} \label{eq:disbulge4}
\log(\sigma/{\rm km\,s}^{-1})= 0.21\log(R_b/{h^{-1}\rm pc})+2.58.
\end{equation}

\section{Predictions from the DLB07 model with galaxy disruption}
\label{sec:disresults}
In order to better understand the changes introduced by the new
physics, we start by plotting the new model predictions leaving the
parameters governing the other physics unchanged from the values in
DLB07. In the next section we will use the techniques introduced in \paperone\ to
find a new best fit, tuning the parameters governing the basic physics
of the model in order to agree with a range of observations.

\subsection{The Luminosity Function}
\label{subsubsec:diskband}

\begin{figure}
\centering
\includegraphics[width=8.4cm]{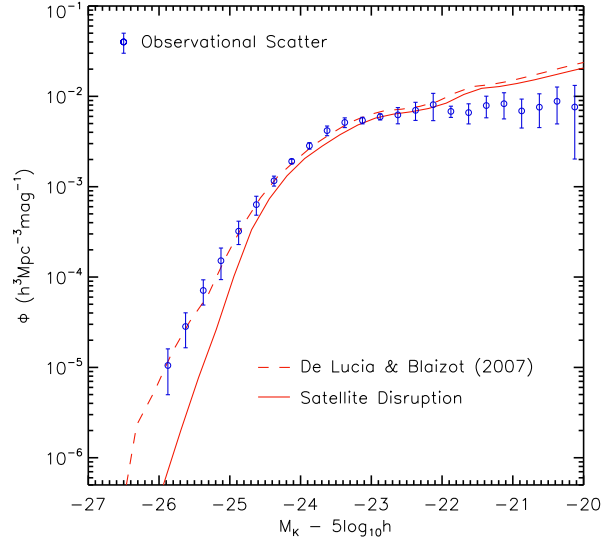}
  \caption{Comparison of the predicted $K$-band luminosity function at
    $z=0$, from DLB07 (dashed red line) and the satellite disruption
    model (solid red line). The data points represent the mean of a
    combined set of observations from \citep{Cole2001, Bell2003,
      Jones2006}, with error bars reflecting the minimum and maximum
    estimates from the three data sets in each bin.}
\label{fig:diskband}
\end{figure}

In Fig. \ref{fig:diskband} we plot the $K$-band luminosity function for the
new model with satellite disruption. The predictions at redshift zero
are compared with the values from the original DLB07
model and with a combination of three different observational data
sets \citep{Cole2001, Bell2003, Jones2006}, respectively from 2DFGRS,
2MASS and 6DFGRS.  The final data points are given by the average of
the maximum and minimum number density estimates in each magnitude
bin, with errors $\sigma_i$ equal to half the difference between
them. The scatter in the combined observational luminosity function
represents the level of accuracy that we require from the model.

The introduction of satellite disruption into the model has a
considerable effect on the bright end side of the $K$-band luminosity
function.  In the framework of a hierarchically growing Universe,
these objects grow from material received during numerous mergers over
their lifetime.  With the inclusion of satellite disruption, a large
amount of material which would otherwise end up in the brightest
objects at redshift zero is transferred from satellites into the
ICM. The result is an excessively low number density of bright objects
in the $K$-band when compared to observations and the previous model.

The number density of dwarf galaxies is, in its turn, less affected
than we inferred in \citet{Henriques2008}. With the self-consistent
implementation, disruption is no longer an instantaneous and dramatic
process; instead galaxies lose material slowly as they follow their
route into the central galaxy. Moreover, the merger time-scale is
increased because satellites become smaller and less affected by
dynamical friction as they spiral inwards.  Nevertheless, the decrease
in the number density of dwarfs is still statistically significant,
due to the large numbers of these objects contained in the Universe.

We emphasize that the poor fit results from the fact that we used
parameters values tuned for the basic physics, without the inclusion
of satellite disruption. This makes it easier to understand the
changes introduced by the new recipe. In the next section, we will
test if there is a region in parameter space where disruption can
produce an overall better agreement between the predicted and observed
luminosity functions.

\subsection{Galaxy Colours}
\label{subsubsec:disredfrac}

\begin{figure}
\centering
\includegraphics[width=8.4cm]{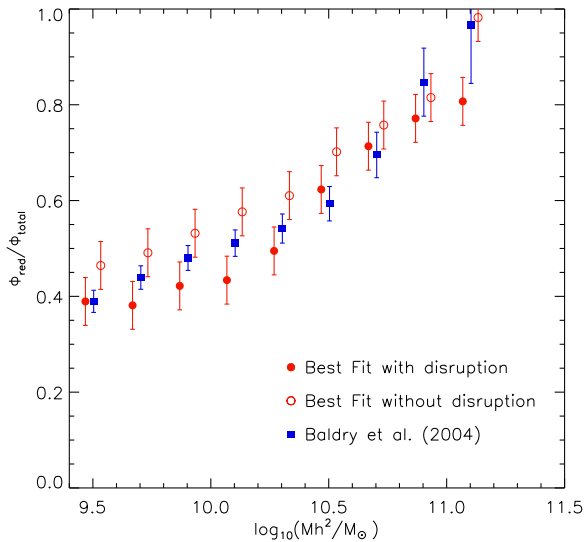}
\caption{The predicted fraction of red galaxies as a function of stellar
  mass. The original \citet{Delucia2007} model (open red circles) is
  compared with the satellite disruption model (filled red circles) and
  observational data from \citet{Baldry2004} (filled blue squares).}
\label{fig:disredfrac}
\end{figure}

In Fig. \ref{fig:disredfrac} we show the fraction of red galaxies as a
function of stellar mass. We divide the galaxies into the two
populations using the selection criteria in \citet{Weinmann2006a},
$(g-r)=0.7-0.032\,(M_r-5\log h+16.5)$, converted into a cut on the
colour-stellar mass relation at redshift zero, $(B-V)=0.065\,\log({\rm
  M}_{\star}h^2/\Msun)+0.09$.  The conversion from the $g-r$ to the
$B-V$ colour was done following \citet{Fukugita1996},
$g-r=1.05(B-V)-0.23$.  The fraction of red galaxies for different mass
bins is then compared with observations from \citet{Baldry2004}.  The
observational masses based on the 'diet' Salpeter IMF \citep{Bell2003}
were reduced by 0.15 dex to agree with the IMF assumed in our SA model
\citep{Chabrier2003}. The predictions from the model with satellite
disruption are compared with the DLB07 model (open red circles) and
observational data from \citet{Baldry2004} (filled blue squares). The
disruption of stellar material from satellites significantly reduces
the fraction of red galaxies over all mass ranges.

This reduction comes about through the removal of predominantly red
stars from galaxies into the ICM.  For central, mostly high-mass
galaxies in halos, this reduces the mass of material accreted from
merging dwarfs. 

\subsection{The Black Hole-Bulge Mass Relation}
\label{subsec:disbhbm}

\begin{figure}
 \centering
 \includegraphics[width=8.4cm]{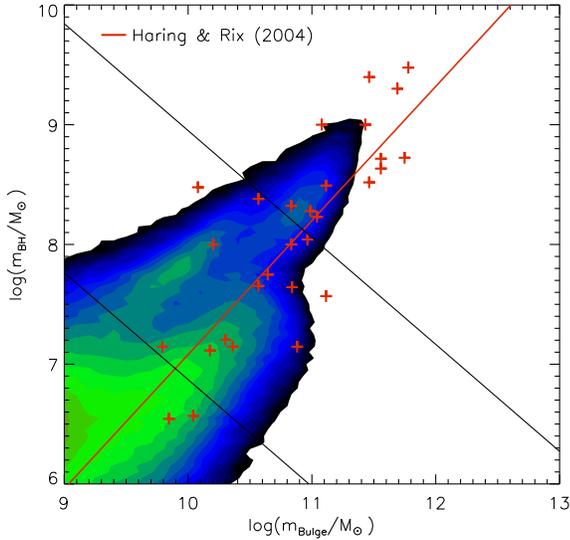}
 \caption{The black hole-bulge mass relation for the satellite
   disruption model (solid contours). The red crosses represent
   observations from \citet{Haring2004} with the best fit to the data
   points given by the red line. The black lines represent the binning
   used to compare the model with observations in sections
   \ref{sec:dismcmc} and \ref{sec:dismcmcbestfit}.}
\label{fig:disbhbm}
\end{figure}

In Fig. \ref{fig:disbhbm} we show the black hole-bulge mass relation
for the satellite disruption model. The colours follow the number
density of objects with blue representing low and green high density
regions.  In comparison with the predictions for the original model
(Fig. 5 in \paperone), there is an overall reduction in the masses of
both bulges and of black holes.

The reduction on black hole masses is determined by the black hole
growth implementation in the model. The build up in black hole mass in
the original DLB07 model is mostly due to the quasar mode:
\begin{equation} \label{eq:msamfbh}
\Delta m_{\rm BH,Q}=\frac{f_{\rm BH}(m_{\rm sat}/m_{\rm
    central})\,m_{\rm cold}}{1+(280\,\mathrm{km\,s}^{-1}/V_{\rm vir})^2}.
\end{equation}
This equation represents the fact that, during a merger event, the
amount of cold gas driven into the central black hole depends on the
instabilities created. The instabilities themselves depend on the mass
ratio between the two merging galaxies. Satellite disruption
decreases the overall mass of satellites, reducing the instabilities
created during mergers and hence the black hole growth due to cold gas
accretion.

Looking at Fig. \ref{fig:disbhbm} we see that the reduction in black
hole growth is more significant than in bulge growth. The peak
previously seen around a bulge mass of $10^{10}\Msun$ and a black hole
mass of $10^{8}\Msun$ has been reduced, with objects moving to lower
black hole masses.

\subsection{Intra-Cluster Light}
\label{subsubsec:disicl}

On of the greatest advantages of introducing stellar disruption into
semi-analytic models is that it naturally explains the production of
intra-cluster material. For the original set of parameters the
disruption model predicts ICL fractions of approximately 22 per cent
for the most massive clusters of galaxies
$M_{\rm{vir}}=10^{15}\Msun$. In section \ref{subsec:icl} we will
compare these predictions with those of an optimized model for virial
masses ranging from $10^{12}$ to $10^{15} \Msun$.



\subsection{The Metallicity of Stars}
\label{subsubsec:dismetals}

\begin{figure}
\centering
\includegraphics[width=8.4cm]{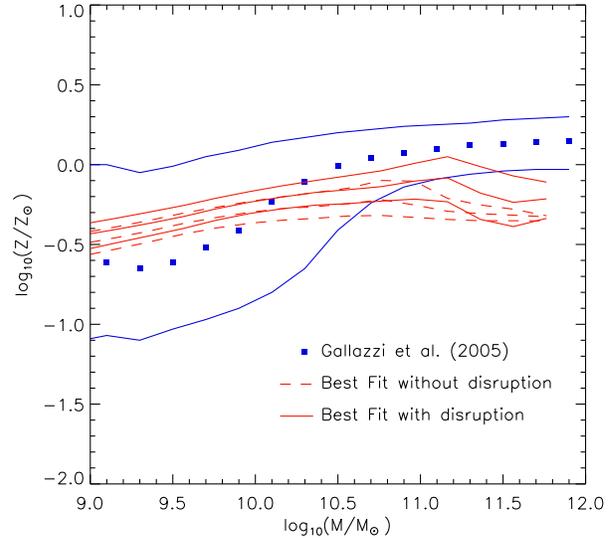}
\caption[Metallicity of stars for the satellite disruption
 model]{Comparison between the metallicity of stars in
  the satellite disruption model (solid red lines), in
  \citet{Delucia2007} (dashed red lines) and in observations from
  \citet{Gallazzi2005} (blue squares and lines). For all the data sets,
  the central line represents the median value of metallicity in each mass
  bin (the blue squares for the observational data), while the upper and
  lower lines represent the 16th and 86th percentiles of the
  distribution.}
\label{fig:dismetals}
\end{figure}

Apart from the galaxy quantities that originally motivated the
introduction of a new recipe into the SA model, there is
another predicted property that is significantly improved by the
disruption of satellite galaxies. In Fig. \ref{fig:dismetals} we show
the metallicity of stars as a function of the galaxy stellar mass. The
new disruption model (solid red lines) is compared with the original
DLB07 predictions (dashed red lines) and with observations from
\citet{Gallazzi2005} (blue squares and lines). The central line
represents the median value of metallicity in each mass bin (the blue
squares for the observational data), while the upper and lower lines
represent the 16th and 86th percentiles of the distribution. From the
figure it is clear that satellite disruption improves the agreement
between SA predictions and observations, by increasing the
metallicity of intermediate and high mass galaxies,
M$_{\star}>10^{10}\Msun$. The explanation for this effect is related
to the decrease of the number density of the most massive
galaxies. With the implementation of disruption, a considerable amount
of material that would otherwise end up in these objects is now
transferred into the ICL. In terms of the metallicity of stars, this
results in massive objects receiving less low metallicity material
from satellites, hence increasing their mean metallicity.

\section{MCMC Parameter Estimation}
\label{sec:dismcmc}
In \paperone\ we have implemented MCMC sampling in the Munich
SA model.  This technique allows us to combine the
constraining power from multiple observational data sets with a fast
sampling of high-dimensional parameter spaces. By doing so we can
verify the level of agreement with observations and the relative
weight of different observations in the final choice of the parameters
in the best fit model, in a statistically-consistent way. Moreover,
whenever reasonable agreement proves to be impossible, it helps us
understanding whether there is a failure in determining the right
parameter configuration, whether there is a fundamental problem with
the underlying model, or whether the introduction of new physics is
required. 

In the previous section we presented the predictions for the
SA model with a self-consistent treatment of the disruption
of satellite galaxies. These were obtained using the parameters from
the original model of DLB07, not adjusted to incorporate this new
physical recipe. In this section, we will use the MCMC sampling
techniques introduced in \paperone\ to search for the combination of
parameters that gives the best fit for the satellite disruption
model. We refer the reader to that paper for a full description of the
statistical methods. The same MCMC implementation is used, namely the
Metropolis-Hastings algorithm \citep{Metropolis1953,Hastings1970} and
a log-normal proposal distribution with a width that assures an
overall acceptance rate in each chain between 10\% and 40\%.

Due to the computational requirements, we will once more perform the
sampling in a single volume of the millennium simulation, representing
1/512 of the total simulation. Since we choose a representative file,
with a luminosity function and mean density similar to that of the
total volume, we are able to correctly constrain the properties of
galaxies in all mass ranges except for the most massive objects
($M_{\star}>10^{11}\Msun$). The MCMC sampling is performed over
$\sim$30 000 steps and the output analyzed using \textsc{getdist},
which is part of the \textsc{cosmomc} software package
\citep{Lewis2002}, adapted to produce 1d and 2d maximum likelihood
(profile) and MCMC marginalised (posterior) distributions.  Once the
best fit parameters are obtained, we then re-run the SA
code over the entire Millennium to obtain the galaxies properties
presented later in this paper.

Using one file representative of the total volume significantly
reduces the computational time required for our study.  Nevertheless,
the size of the calculations involved, even in that smaller volume,
still make it a challenging task. In order to perform the sampling,
the Cosmology Machine (COSMA) supercomputer supplied by Sun
Microsystems was used, a machine based in the Durham University that
is part of the Virgo Consortium facilities.

\subsection{Model Parameters}
\label{subsec:parameters}
As in \paperone, in order to better understand the basic physics of the
model, we choose to do our MCMC sampling only on 6 of 12 parameters in
the model.  On the top line of table \ref{table:hybridoursampar} we
show the frozen parameters corresponding to the baryon fraction
($f_{\rm b}$), the redshifts of beginning and end of reionization
(respectively $z_0$ and $z_{\rm r}$), the major to minor merger
threshold ($T_{\rm merger}$), the instantaneous recycled fraction
($R$) and the yield of metals ($Y$). On the bottom line, we show the
values for the more fundamental parameters that we choose to
sample. These are the star formation efficiency ($\alpha_{\rm SF}$),
the AGN radio mode efficiency ($k_{\rm AGN}$), the black hole growth
efficiency ($f_{\rm{BH}}$), the SN reheating and ejection efficiency
(respectively $\epsilon_{\rm{disk}}$ and $\epsilon_{\rm{halo}}$) and
ejected gas reincorporation efficiency ($\gamma_{\rm ej}$). The
physics governed by all the parameters are fully described in
\citet{Croton2006} as well as in \paperone.

\begin{table}
\begin{center}
  \caption{SA model parameters from
    \citet{Delucia2007}. The first 6 parameters are frozen in our
    analysis at the values shown here.}
\label{table:hybridoursampar}
\begin{tabular}{cccccc}
\\[3pt]
\hline
\hline
$f_{\rm b}$& $z_0$& $z_{\rm r}$& $T_{\rm merger}$& $R$& $Y$\\
0.17& 8& 7& 0.3& 0.43& 0.03\\
\hline
\hline
$\alpha_{\rm SF}$ &$k_{\rm AGN}$ &$f_{\rm{BH}}$ &$\epsilon_{\rm{disk}}$
&$\epsilon_{\rm{halo}}$ &$\gamma_{\rm ej}$\\
0.03& $7.5\times10^{-6}$& 0.03& 3.5& 0.35& 0.5\\
\hline
\hline
\end{tabular}
\end{center}
\end{table}

\subsection{Observational Constraints}
\label{subsec:obsconstraints}

The comparison between model and observations will be done using the
same data sets and statistical tests as before and since we have not
introduced any additional parameter, the relative goodness of the
original and the satellite disruption model can be assessed
directly. The ability or not for the MCMC to find a best fit model
with a likelihood higher than before will tell us if the introduction
of the new physical recipe is justified.

The three observations used are the $K$-band luminosity
function, the fraction of red galaxies versus stellar mass and the
black hole-bulge mass relation.  For a detailed description of the
observational data sets and the statistical tests employed, we refer
the reader to \paperone.

\section{Results from the optimised stripping model}
\label{sec:dismcmcbestfit}
\subsection{The Best Fit Semi-Analytic Model with Satellite Disruption}

We now present the results for the MCMC sampling when the
SA model with satellite disruption is constrained by the
three observational data sets combined. The likelihood for the best
fit model is given by the product of the likelihood from the
three statistical tests:
\begin{equation} \label{eq:disruptionliketotal}
\pi(x_i)=\mathcal{L}_{(\rm K-band)}\times\mathcal{L}_{(\rm Colour)}
\times \mathcal{L}_{(\mathrm{BH-Bulge})}.
\end{equation}  

\begin{figure}
\centering
\includegraphics[width=8.4cm]{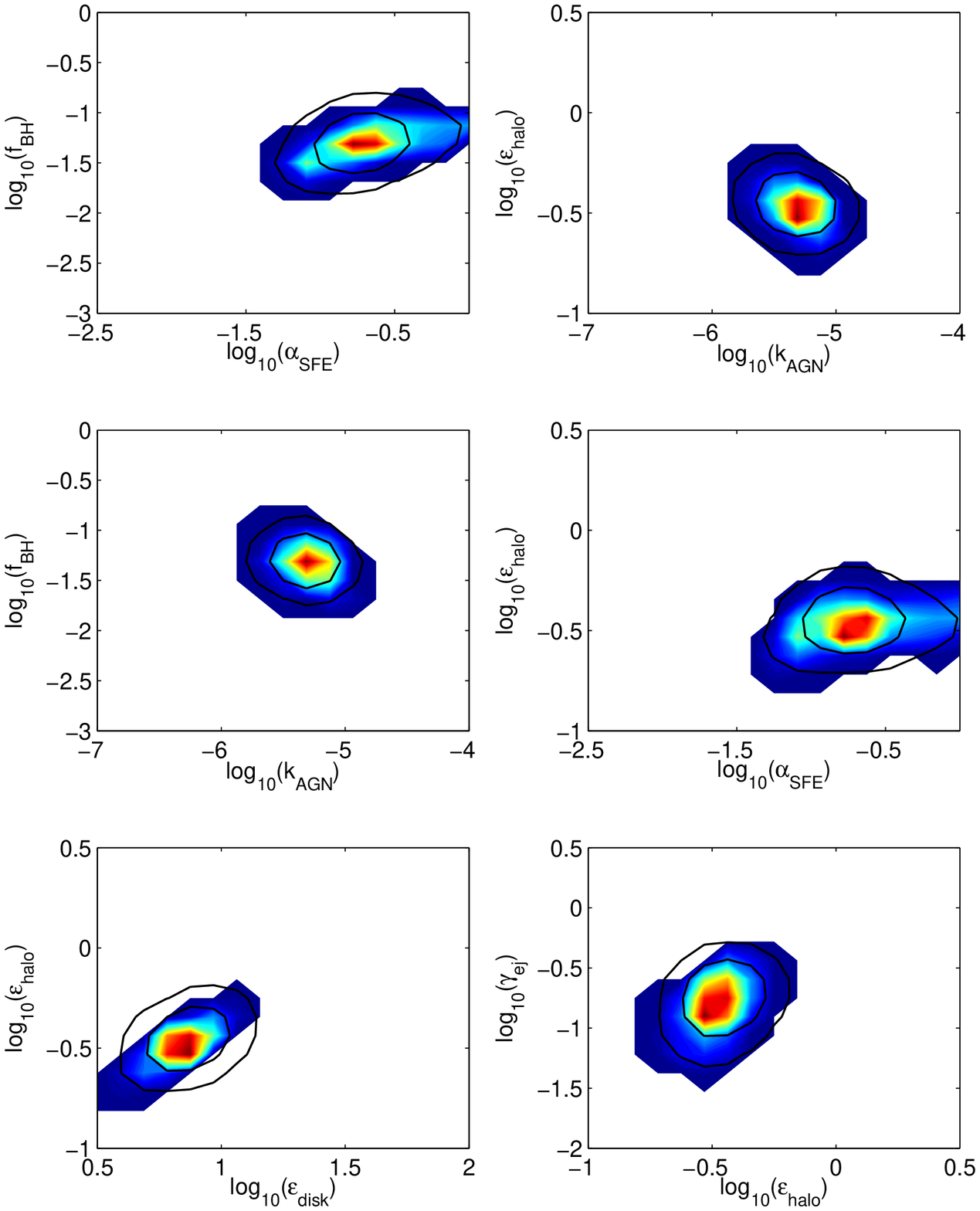}
\includegraphics[width=8.4cm]{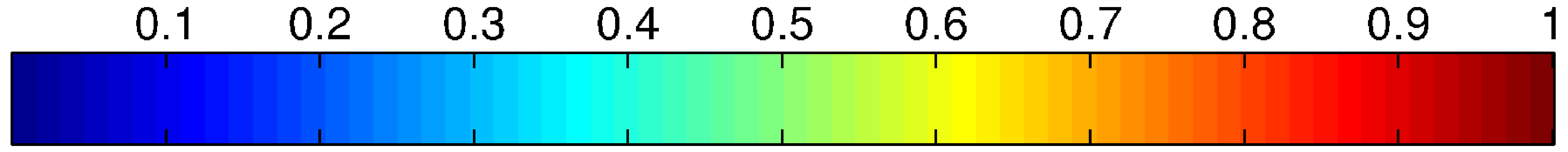}
\caption{Correlations between the 6 parameters analysed in the study
  for the SA model with satellite disruption constrained by
  three observational properties: the $K$-band luminosity function,
  the fraction of red galaxies, and the black hole-bulge mass
  relation. The values of the parameters are plotted in log space,
  with the solid contours representing the 68\% and 95\% preferred
  regions from the MCMC (the posterior distribution) and the colours
  the maximum likelihood value sampled in each bin (the profile
  distribution). The colour scale is normalized by the maximum
  likelihood value of 0.15.}
\label{fig:dismcmc2d}
\end{figure}

Fig. \ref{fig:dismcmc2d} shows the allowed ranges and correlations
between the parameters sampled. As in \paperone, the use of combined
observations to produce one comprehensive data set in terms of galaxy
formation properties restricts the parameters to one small region
with acceptable likelihood.

The original model had a maximum likelihood value of just 0.037 when
compared with the chosen set of observational constraints, meaning it
was formally ruled out at a 2-$\sigma$ level. However, the new
satellite disruption recipe brings the model likelihood up to
0.15. Since we introduced no additional parameters, this increase in a
factor of 4 in peak likelihood means that the inclusion of the new
physical process seems to be favoured by data.

In comparison to our analysis for the model without galaxy disruption,
we see both an increase in the likelihood value of our best fit and in
the allowed regions in parameters space. This is mainly caused by the
changes that the new physical recipe produces on the black hole-bulge
mass relation. The regions required in parameter space by the
different tests now have a much larger overlap. This means that the
black hole and bulge mass build up in the new model is now consistent
with the other galaxy properties analyzed, namely the $K$-band
luminosity function and the fraction of red galaxies.

In order to understand the changes introduced by satellite disruption
we include two additional plots in our analysis (the two top panels in
Fig. \ref{fig:dismcmc2d}), the correlations between the AGN quasar
mode parameter and the star formation efficiency, and the AGN radio
mode and the disk reheating efficiency.

An expected correlation, but one that was not seen in \paperone\, is
shown in the upper left panel. A positive correlation is now evident
between the quasar mode, responsible for the black hole growth, and
the star formation efficiency, responsible for the bulge
growth. Considering the physics governed by them, it comes at no
surprise that an increase in one requires an increase in the other in
order to maintain the black hole-bulge mass fraction of galaxies. The
fact that this correlation was not present in the previous study is
explained by the considerably smaller regions with acceptable
likelihood that we found before.

For the same reason, the correlation between the supernova reheating
and ejection parameters, produced by the $K$-band luminosity
constraint, is now clear in the lower left corner. As explained in
\paperone\ this correlation keeps the virial velocity cutoff, above
which SN feedback stops being effective, at a constant value:
\begin{equation} \label{eq:hybridejectvvir}
V_{\rm vir,0}=\left(\epsilon_{\rm halo}\over\epsilon_{\rm
  disk}\right)^{1\over2}\,V_{\rm SN}.
\end{equation}
This means that this form of energy only suppresses star formation in
the smallest objects where an excess compared to observations was
previously seen.

Accordingly, the lower and middle right panels show an identical
behaviour for the parameters as in the $K$-band sampling on
\paperone. An increase in the amount of gas ejected by SN needs to be
balanced by an increase in the reincorporation time-scale and the star
formation efficiency.

\subsection{Best Fit Parameters and Confidence Limits}
\label{subsec:disbestfitparams}

\begin{table*}
\begin{center}
  \caption{Statistics from the MCMC parameter estimation for the
    parameters in the satellite disruption model. The best fit and
    marginalized confidence limits are compared with the published
    values from \citet{Delucia2007} and with the best fit values
    obtained without the inclusion of satellite disruption
    (\paperone).}
\label{table:dismargestats}
\scalebox{0.84}{%
\begin{tabular}{lccccccc}
  \\[3pt]
  \hline
  \hline
  &DLB07 &\paperone  &Disruption Model &-2$\sigma$  &-1$\sigma$  &+1$\sigma$ &+2$\sigma$\\
  \hline
  \hline
  $\alpha_{\rm{SF}}$ &0.03 &0.039 &0.17 &0.078 &0.13 &0.28 &0.53\\
  \hline
  $k_{\rm{AGN}}$ &$7.5\times10^{-6}$ &$5.0\times10^{-6}$& $5.3\times10^{-6}$ &$2.7\times10^{-6}$ &$3.7\times10^{-6}$ &$6.2\times10^{-6}$&$7.9\times10^{-6}$\\
  $f_{\rm{BH}}$ &0.03 &0.032 &0.047 &0.030 &0.041 &0.061 &0.075 \\ 
  \hline
  $\epsilon_{\rm{disk}}$ &3.5 &10.28 &6.86 &5.22 &6.33 &8.51 &10.11 \\
  $\epsilon_{\rm{halo}}$ &0.35 &0.53 &0.33 &0.26 &0.31 &0.40 &0.46\\
  $\gamma_{\rm{ej}}$ &0.5 &0.42 &0.13 &0.076 &0.12 &0.24 &0.30 \\
  \hline
  \hline
\end{tabular}}
\end{center}
\end{table*}

The best fit and confidence limits for the 6 free parameters in the
model with satellite disruption, together with the published values
from DLB07 and the best fit for the model without satellite disruption
are shown in table \ref{table:dismargestats}.  Despite a different
model being used, the parameters from both the original DLB07 and the
previous MCMC analysis remain close to the new 2$\sigma$ confidence
limits.

In comparison to the best fit obtained for the model without satellite
disruption, the most dramatic effect in terms of preferred parameter
values is the increase in the star formation efficiency. It represents
a change in the fraction of cold gas transformed into stars in a disk
dynamical time from 4\% to 16\%. Such an increase is required to
repopulate the bright end side of the $K$-band luminosity function,
largely affected by disruption. This effect is somewhat balanced by an
increase in the mechanical heating from AGN, given by the product of
$k_{\mathrm{AGN}}$ and $f_{\mathrm{BH}}$.

The virial velocity cutoff, above which supernova stops being
effective (Eq. \ref{eq:hybridejectvvir}), is similar to that found for
the model without disruption. This means that the new model still
requires the supernova feedback to be effective only for small
objects. However, the strength required from this form of feedback is
now smaller (6.94 for the reheating and 0.33 for the ejection
efficiency), which seems to be in better agreement with observations
\citep{Martin1999}.  This means that disruption does indeed help
reducing the excess of dwarfs in the model and that an excessively strong
feedback is no longer required.

\subsection{Predictions for the Best Fit Satellite Disruption Model}
\label{subsec:dismcmcbestfitpredictions}

In this section we analyze the predictions from the best fit model
with satellite disruption. Contrary to the MCMC sampling, done in a
single volume due to computational resource limitations, the following
predictions were obtained using the full Millennium Simulation.

\subsection{Galaxy Luminosity Functions}
\begin{figure*}
\centering
\includegraphics[width=8.4cm]{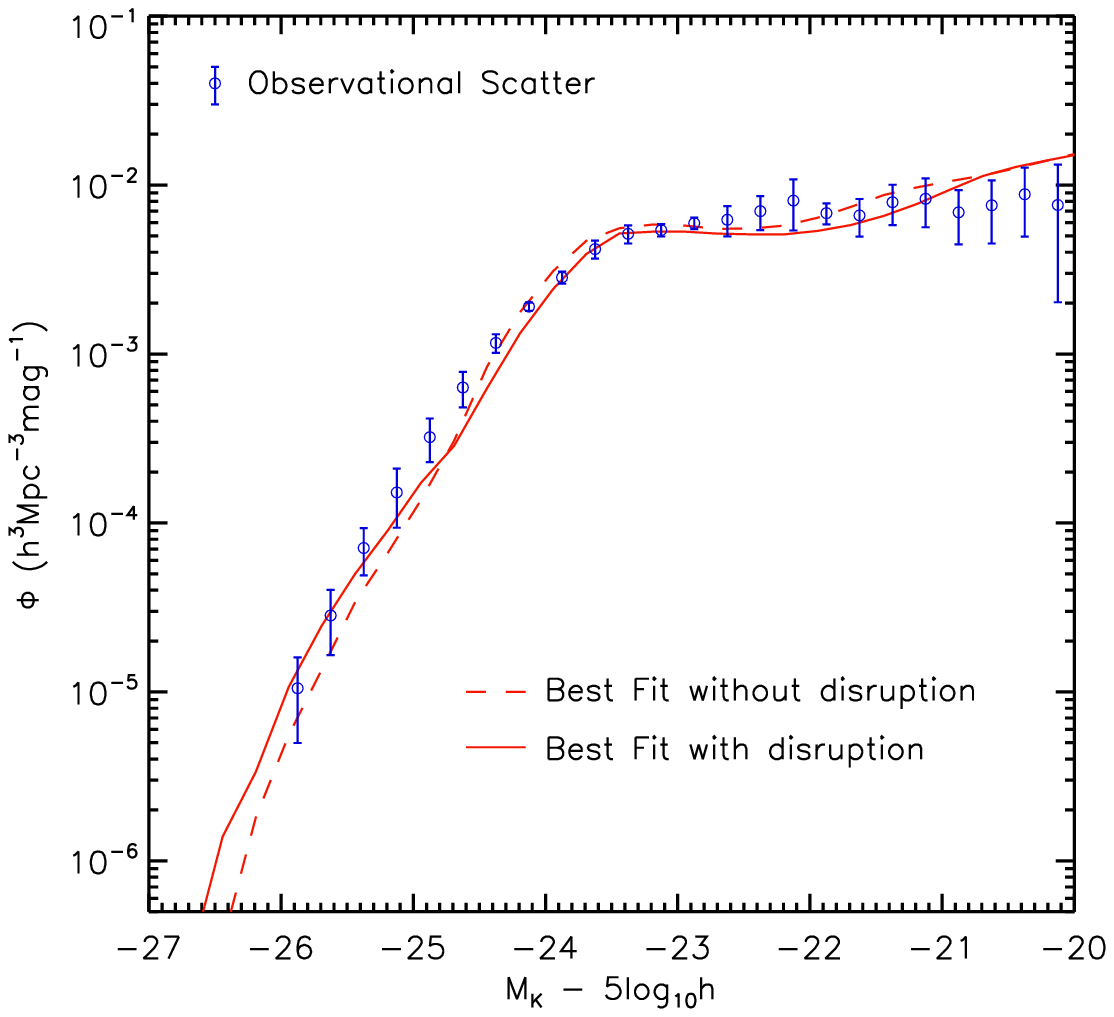}
\includegraphics[width=8.4cm]{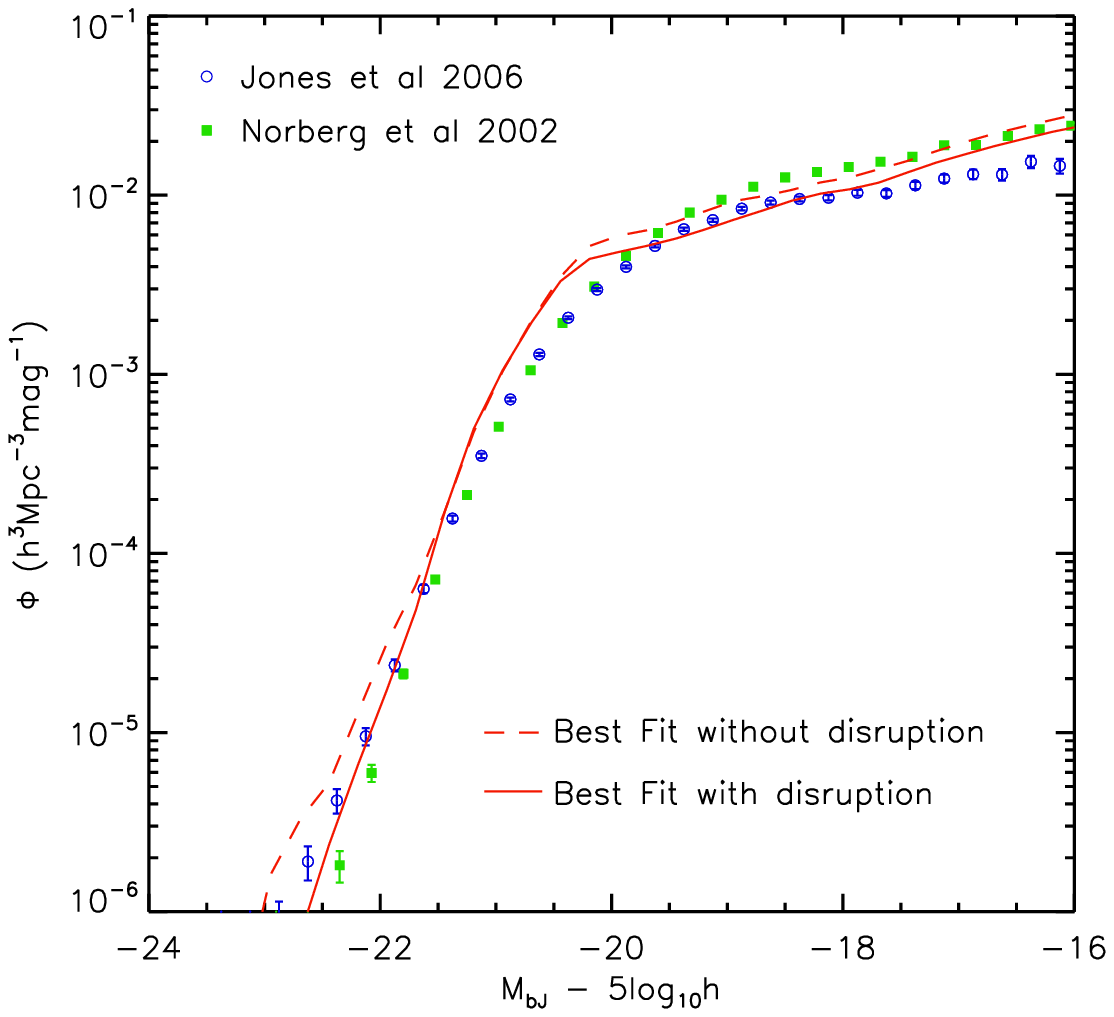}
\caption{Comparison of the predicted $K$-band (left panel) and
  \bj-band (right panel) luminosity functions at $z=0$ from the best
  fit for DLB07 (dashed red line) and the best fit for the satellite
  disruption model (solid red line). On the left panel, the data
  points represent the observations used to constrain the luminosities
  of galaxies in the MCMC parameter estimation \citep{Cole2001,
    Bell2003, Jones2006}. On the right panel, the \bj-band luminosity
  function is compared with observations from 2DFGRS (green filled
  squares) and 6DFGS (blue open circles), respectively
  \citet{Norberg2002} and \citet{Jones2006}.}
\label{fig:disbestfitkband}
\end{figure*}

In the left panel of Fig. \ref{fig:disbestfitkband} the $K$-band
luminosity functions from the best fit for DLB07 and the best fit for
the satellite disruption models are plotted against the observational
data set used to constrain the sampling.  In comparison to the best
fit obtained for a model without disruption the new model produces
similar galaxy luminosities. However, for the model including
satellite disruption, the good agreement at the low luminosity end is
achieved with a less efficient supernova feedback, which seems to
be in better agreement with observations.

In the right panel of Fig. \ref{fig:disbestfitkband} we show the
predictions for the \bj-band luminosity function. The best fit
satellite disruption model is compared with the best fit for DLB07 and
two observational data sets.  As for the $K$-band, both best fit
models with and without disruption show produce similar predictions,
achieving an overall good agreement with observations, except for the
region around $L_*$. We note however that the \bj-band flux is highly
dependent on the adopted dust model (which does not happen with the
other properties analyzed). For this reason, the excess just mentioned
can be removed by adjusting the dust implementation, without affecting
the agreement achieved for the other galaxy properties. We choose not
to do so, in order to make more clear the changes introduced by
satellite disruption.

\subsection{Galaxy Colours}
\begin{figure}
\centering
\includegraphics[width=8.4cm]{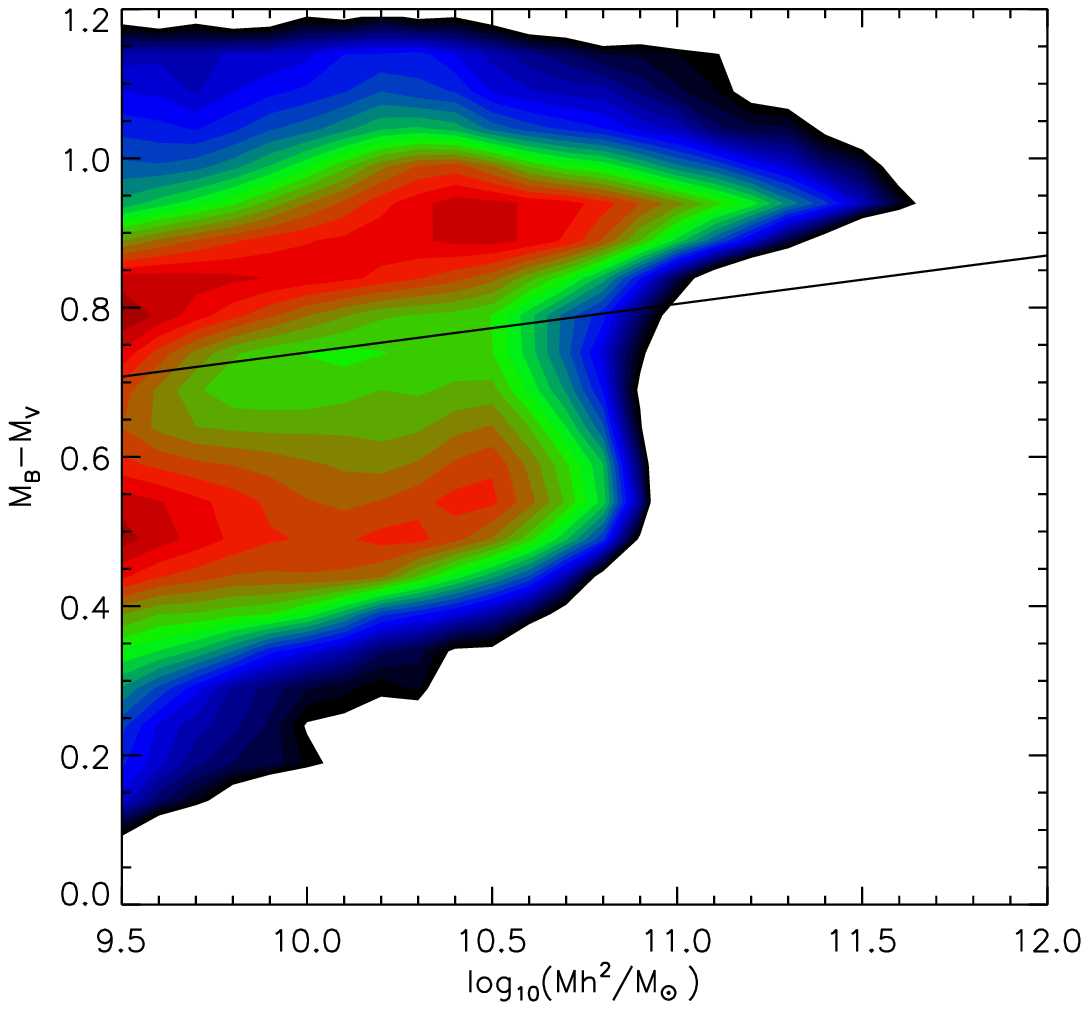}
\includegraphics[width=8.4cm]{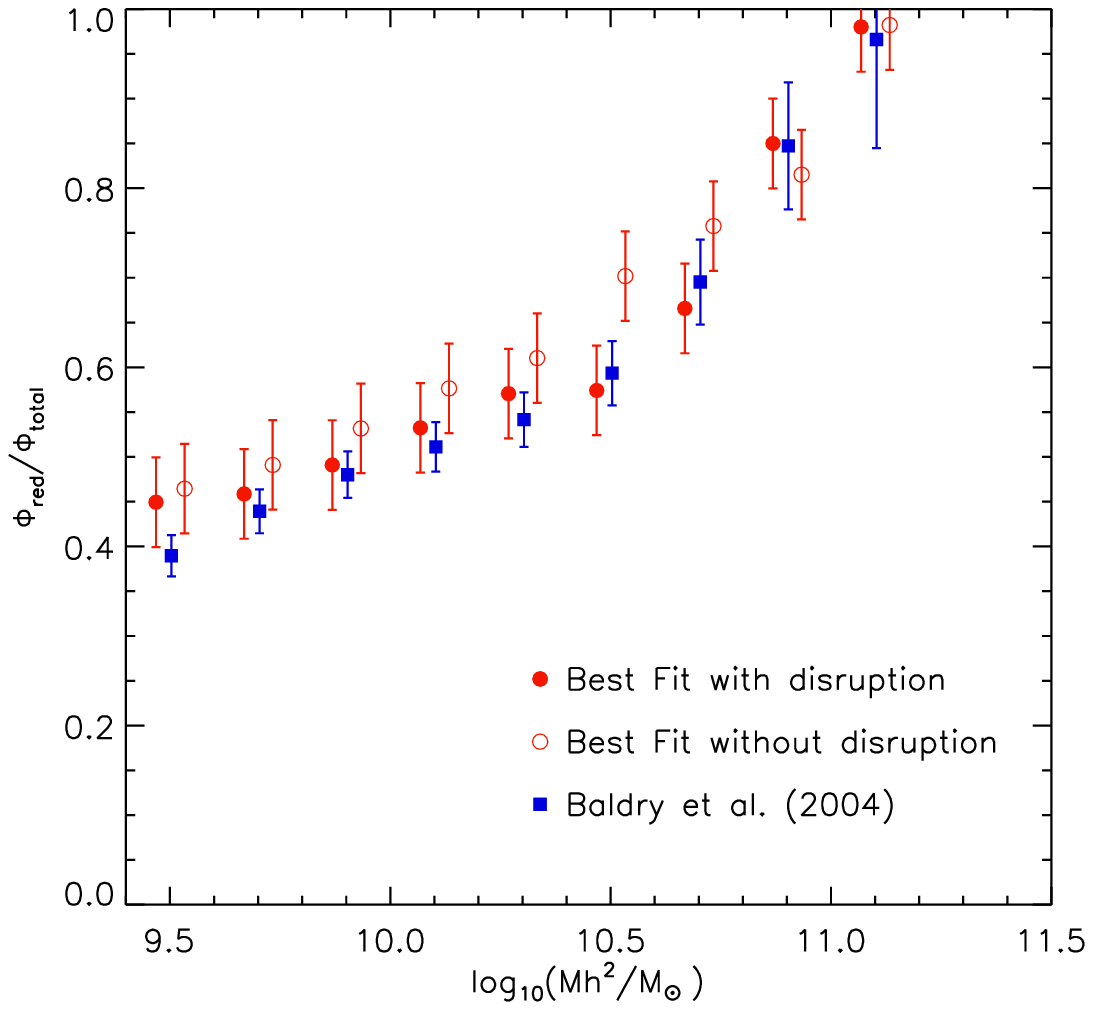}
\caption{The top panel shows the $B$-$V$ colour-stellar mass relation
  for the galaxies in the best fit satellite disruption model. The
  solid line represents the division between the red and blue
  populations in \citet{Weinmann2006a}. The predicted fraction of red
  galaxies as a function of stellar mass is showed in the bottom
  panel. The best fit for the original DLB07 model (open red circles)
  is compared with the best fit for the satellite disruption model
  (filled red circles) and observational data from \citet{Baldry2004}
  (filled blue squares).}
\label{fig:disbestfitcolorfract}
\end{figure}

Fig. \ref{fig:disbestfitcolorfract} shows the predictions for the
galaxy colours in the best fit model with satellite disruption. The
top panel gives the $B$-$V$ colour-stellar mass relation (colour coded
by the number density of objects), while the bottom panel shows the
fraction of red over the total number of galaxies as a function of
stellar mass. The introduction of disruption without changing the
original DLB07 parameter values decreases the overall fraction of red
galaxies, causing the model to under-predict the number of these
objects except for L$_{\star}$ galaxies
(Fig. \ref{fig:disredfrac}). As seen in the bottom panel of
Fig. \ref{fig:disbestfitcolorfract} , the MCMC optimization brings the
model into agreement with observations for the entire plotted range.
When compared to the best fit without disruption, the new model
produces an overall better agreement with observations, reducing the
systematic excess of small galaxies previously seen (open red
circles).

More interesting is that the model now starts to produce an isolated
population of massive red galaxies, as required by observations (top
panel of Fig. \ref{fig:disbestfitcolorfract}). However, despite
reducing the number of dwarf red galaxies, the disruption model still
has an excessive fraction of these objects when compared to
observations.

\subsection{The Black Hole-Bulge Mass Relation}
\label{subsec:disbestfitbhbm}

The changes introduced by satellite disruption in the predicted
black-hole bulge mass relation are the main reason for the better fit
obtained in the new model. This means that the allowed likelihood
region for the $K$-band luminosity function and for the fraction of
red galaxies now produces black hole and bulge masses that agree with
observational results.

As explained in section \ref {subsec:disbhbm}, disruption has a large
impact on both black hole and bulge masses. However, looking at the
distribution of these masses in the best fit model, it becomes clear
that the higher likelihood obtained is largely due to the reduction in
black hole growth. The small sizes of the satellites will mean that
smaller instabilities are created during mergers and that smaller
amounts of cold gas are driven into the central black holes.

\begin{figure}
 \centering
 \includegraphics[width=8.4cm]{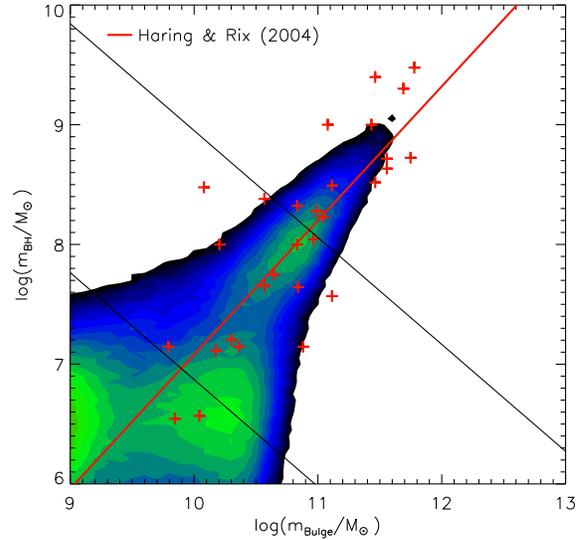}
 \caption{The black hole-bulge mass relation for the best fit
   satellite disruption model (solid contours). The red crosses
   represent observations from \citet{Haring2004} with the best fit to
   the data points given by the red line. The black lines represent
   the binning used to compare the model with observations.}
\label{fig:disbestfitbhbm}
\end{figure}

In Fig. \ref{fig:disbestfitbhbm} we show the black hole-bulge mass
relation for the satellite disruption model with the best fit
parameters. The peak around bulge masses of $10^{10}\Msun$ and black
hole masses of $10^{8}\Msun$ has disappeared. This means that most of
the intermediate mass bulges and black holes are now on or below the
observational red line, where most observational points are
located. The binomial test used produces a probability three times
higher than before.

\subsection{The Intra-Cluster Light}
\label{subsec:icl}

\begin{figure}
\centering
\includegraphics[width=8.4cm]{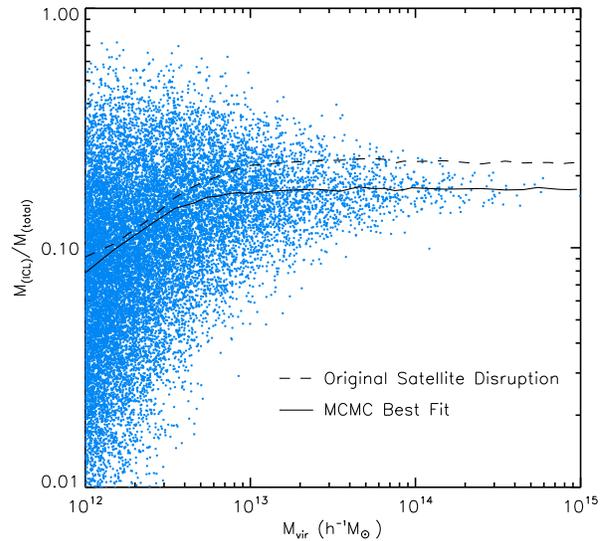}
\caption{Fraction of the mass in the ICM over the
  total stellar mass of the group as a function of virial mass. The
  blue dots are a representative sample of the total galaxy population
  in the best fit model with satellite disruption. The solid and
  dashed lines represent the median of the $M_{\rm
    {ICL}}$/$M_{\rm{total}}$ distribution for the satellite disruption
  model with the best fit and original parameters respectively.}
\label{fig:disbestfiticl}
\end{figure}

In Fig. \ref{fig:disbestfiticl} we plot the predicted ICL
from our best fit model with satellite disruption. The solid line
represents the median of the $M_{\rm {ICL}}$/$M_{\rm{total}}$
distribution in each bin while the dashed line gives the same
relation for the disruption model with the original set of parameters. 

The predicted ICL fraction for groups with $M_{\rm{vir}}>10^{13}\Msun$
has a mean of $\approx 18\%$. This value is compatible with
observations that detect ICL fractions between 10\% and 40\%
\citep{Bernstein1995, Gonzalez2000, Feldmeier2002, Feldmeier2004b,
  Gonzalez2005, Zibetti2005}. The prediction decreases systematically
has we move to smaller virial masses reaching a mean of $\approx 7\%$
for $M_{\rm{vir}}=10^{12}\Msun$. The difficulty in distinguishing
between ICL and that from the halo of the central galaxy (the former
can be regarded as an extension of the latter), means that
observational data is only available for large groups and has a
considerable scatter.

\subsection{The Metallicity of Stars}
\begin{figure}
\centering
\includegraphics[width=8.4cm]{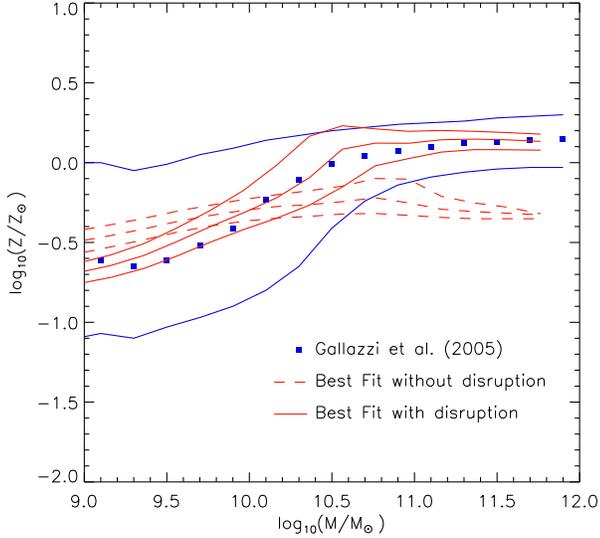}
\caption{Comparison between the metallicity stars in the best fit for
  the satellite disruption model (solid red lines), in the best fit
  for DLB07 (dashed red lines) and in observations from
  \citet{Gallazzi2005} (blue squares and lines). For all the data
  sets, the central line represents the median value of metallicity in
  each mass bin (the blue squares for the observational data), while
  the upper and lower lines represent the 16th and 86th percentiles of
  the distribution.}
\label{fig:disbestfitmetals}
\end{figure}

In Fig. \ref{fig:disbestfitmetals} we show the metallicity of stars as
a function of the galaxy stellar mass.  We remind the reader that this
was not one of the observations used to constrain the model.  The best
fit disruption model (solid red lines) is compared with the best fit
for the original DLB07 (dashed red lines) and with observations from
\citet{Gallazzi2005} (blue squares and lines). As in
Fig. \ref{fig:dismetals}, the central line represents the median value
of metallicity in each mass bin (the blue squares for the
observational data), while the upper and lower lines represent the
16th and 86th percentiles of the distribution.

As for the best fit model without disruption, the increase in
supernova feedback required by the MCMC lowers the metallicity of
stars for the low mass galaxies in the best fit for the disruption
model. In comparison to the original model (Fig. \ref{fig:dismetals}),
for this mass range, the higher SN feedback from both best fit models
increases the fraction of metals in the gas phase, consequentially
lowering the fraction of metals in stars.

As explained in section \ref{subsubsec:dismetals}, the introduction of
disruption increases the metallicity of massive objects, by reducing
the amount of low metallicity material that they receive from
mergers. This means that the distribution of metals in the model is
now in close agreement with observations over the entire mass range
plotted.

\section{Discussion}
\label{sec:dismcmcconclusions}

In \paperone, we introduced a new approach to galaxy formation
modelling. Combining the MCMC sampling techniques with SA models
allowed us to gain insight on the physical importance of the different
galaxy parameters.  With the introduction of a new physical recipe,
presented here, both the normalisation and shapes of the acceptable
likelihood regions contain useful information that can be used to
discriminate between models.

In this work we implement a new physical ingredient in the
SA recipe: the stripping of satellite galaxies by tidal
forces during merging events. The concept had already been introduced
in \citet{Henriques2008}, but only as an \emph{a posteriori}
study. That approach, despite enabling us to gain insight on the
impact that this process would have in the population of dwarfs,
did not allow us to study the effect of stripping in a self-consistent
way, and in particular the effect on the most massive galaxies.

The self-consistent implementation that we describe in this chapter
makes it possible to study the impact of disruption on the properties
of galaxies of all types. Namely, it allows us to study the loss in
mass by satellite galaxies, the slower build up of central galaxies
(since they receive less material from satellites) and the growth of
the ICM component, previously neglected. On top of that, the MCMC
sampling allows us to learn if the new process is favoured or not by
observations.

In comparison to the original model, the best fit likelihood of the
model with disruption is \changed{four times} higher with respect to
the $K$-band luminosity Function, the fraction of red galaxies and the
black hole-bulge mass relation.

\begin{itemize}
\item Since we did not introduce any additional parameter to model
  disruption, this means that the inclusion of tidal disruption of
  stellar material from satellite galaxies during mergers seems to be
  favoured by observations.
\item Moreover, it means that the new model is now formally
  consistent with the combined observational data set we used.
\item The higher likelihood value and the larger allowed
    likelihood regions in parameter space of the new best fit model
    are mainly determined by the changes produced by satellite
    disruption on the black hole-bulge mass relation. This means that
    the growth of bulges and black holes is now consistent with the
    properties of the galaxies as a whole.
\item The new best fit has a considerably higher star formation
  efficiency in order to correct the reduction in the number density
  of massive objects caused by disruption. This is balanced in massive
  galaxies by an increase in the AGN feedback efficiency. At low
  masses, a less effective supernova feedback is now required, in
  better agreement with observations \citep{Martin1999}.
\item Although they were not used as constraints, the MCMC sampling
  for the new model kept the agreement with the observational \bj-band
  luminosity function and significantly improved the overall shape of the
  metallicity distribution of stars.
\item Finally, the introduction of disruption allows us to follow the
  build up of the ICM, which represents about 18\% of the total light in
  clusters, in agreement with observations (between 10\% and 40\%).
\end{itemize}

\subsection{Future Challenges}
\label{dismcmcconclusionsincon}

Despite the higher likelihood found for the best fit model,
SA predictions are still far from exactly reproducing
observations. From our analysis, the biggest challenge for the
SA model is still to reproduce the properties of dwarf
galaxies. With the introduction of satellite disruption, the model is
able to reproduce the number density of these objects with a less
efficient supernova feedback, which is in better agreement with
observational studies \citep{Martin1999}. 

Nevertheless, this type of galaxies remain predominantly red, in
disagreement with observational studies. Stripping of stars does help
reducing the number of red dwarfs, but is only part of the solution to
the problem.  Another important factor might be the treatment of the
impact of mergers on the gas phase. For example, \citet{Font2008}
consider a model in which the gas of satellite galaxies is
continually, instead of instantaneously stripped, enabling them to
form stars and remain on the blue sequence for longer.

The higher likelihood found for the best fit model with satellite
disruption seems to be directly related to a better agreement between
the black hole-bulge mass and the other two constraints.  This fact is
apparently related to the slower build up of black holes relative to
bulges and the overall slower build up of both components.  However,
we emphasize the points raised in \paperone\ about the simplicity of
the black hole growth model, which still appears to neglect important
features such as the impact of the quasar mode feedback on galaxy
properties. In addition, from an observational point of view, larger
and more robust data sets of black hole properties are needed.

The predictions obtained for the ICL are in good agreement with
observations. However, the available data is still subjected to large
error bars, due to the difficulties in distinguish between the light
from the brightest cluster galaxies and the ICM itself. Also, this
component is still only detectable in large groups and clusters due to
its low surface brightness. If future improvements in observations
make it possible, this property could be used to directly constrain
the model with satellite disruption.

\section*{Acknowledgements}
We thank all the members of the Sussex Survey Science Centre which
joint expertise helped developing the innovative idea in this
paper. We are grateful to Gabriella De Lucia, Guinevere Kauffmann,
Volker Spingel and Simon White for providing us with the Munich SA
code and for supporting our use of it.

The computations developed for this work, were performed in the Virgo
Consortium cluster of computers, COSMA. The authors would like to
thank Lydia Heck for its great technical knowledge about COSMA and
constant feedback without which this work could not have been done. 

BH acknowledges the support of his PhD scholarship from the Portuguese
Science and Technology Foundation which supported him for most of the
time while this work was developed.  PAT was supported by an STFC
rolling grant.

\bibliographystyle{mn2e}
\bibliography{paper}

\label{lastpage}

\end{document}